# 大语言模型和采样边缘化驱动的基于音素的语音识别

马 特[1]，李男杰[1]，黄 浩[1]，欧智坚[2]

（1．新疆大学计算机科学与技术学院，乌鲁木齐 830046；

2．清华大学语音与智能（SPMI）实验室，北京 100084）

**摘 要**：最近，基于大语言模型的音素转文字方法（Large Language Model-based Phoneme-to-Grapheme, LLM-P2G）在语音识别任务中展现出优异性能，成为替代传统 WFST 解码方法的可行方向。该框架通过音素预测与文字生成的两阶段建模，兼顾识别精度与系统扩展性。然而，现有 LLM-P2G 采用 Top-K 边缘化（Top-K Marginalized, TKM）训练策略，其候选音素序列依赖束搜索生成，存在路径多样性不足、训练效率低、资源开销大等问题。为此，本文提出采样边缘化训练策略（Sampling-K Marginalized, SKM），以随机采样替代束搜索生成候选路径，改进了边缘化建模与训练效率。在波兰语与德语数据集上进行实验，结果表明 SKM 在保持模型复杂度的前提下，进一步提升了模型学习收敛速度和识别性能。与利用投射器结合大语言模型的语音识别方法（SpeechLLM）的对比实验亦显示，SKM 驱动的 LLM-P2G 在识别精度和结构简洁性方面更具优势，研究验证了该方法在跨语言语音识别系统中的实用价值与应用潜力。

端到端语音识别模型因其简化的训练流程和良好的识别效果，已成为近年来的研究热点。然而，传统模型通常无法充分利用大语言模型（Large Language Model, LLM）[1-3]强大的语言理解和生成能力，且在多语言语音识别任务中表现受限。这主要是因为端到端模型将声学与语言建模融合在一个黑盒中，限制了模型的泛化和扩展能力。

为解决上述问题，研究者尝试将大语言模型引入语音识别系统，提升识别性能和多语言适应性。现有将大语言模型与语音识别结合的方法主要分为三种：一是大语言模型后处理方法[4]，即先通过传统声学模型生成候选文本，再利用大语言模型进行文本级的后处理和纠错；二是投射器（projector）连接方法[5]，在声学模型与大语言模型之间插入一个神经网络模块，实现从声学神经网络的输出特征向量投射接入语言模型，再进行微调训练；三是结合音素建模与大语言模型的文字转写方法（Large Language Model-based Phoneme-to-Grapheme, LLM-P2G）[6]；该方法以音素作为连接语音与语言的接口，首先由声学模型预测音素序列，随后由大语言模型将其转换为最终文本，属于两阶段结构[7-8]。

第一种后处理方法将语音识别成文本后再进行纠错，显得笨重，系统臃肿，识别性能提升也有限。第二种投射器连接方法引入了额外模块，增加了系统复杂度和训练难度，不利于模型的灵活部署，也很难保证投射精度。相比之下，LLM-P2G 方法通过明确的音素接口，以音素连接语音与语言，音素作为离散符号，可充分直接利用 LLM 强大的语言能力。此外，该方法还引入多候选音素序列训练与解码策略——Top-K 边缘化（Top-K Marginalized, TKM）[6]，有效提升了识别性能，成功取代了传统加权有限状态转换器（Weighted Finite-State Transducer，WFST）[9]解码流程，实现了多语言扩展和灵活解码。

尽管如此，LLM-P2G 模型中的 TKM 策略在候选音素序列生成环节仍依赖束搜索（beam search），该策略存在诸多不足：首先，束搜索容易导致候选路径多样性不足，影响了边缘化概率估计；其次，束搜索的束宽增大时计算资源消耗显著，降低了训练与推理效率；此外，束搜索过程的贪婪性质也可能影响训练的稳定性和收敛速度。为解决这些瓶颈，

**收稿日期**：2025-07-02

**基金项目**：广西科技计划项目（2022AC16002）；国家自然科学基金（62466055）

**作者简介**：马特（1998.12.1），男，硕士研究生

**通信作者**：欧智坚，教授，E-mail: ozj@tsinghua.edu.cn

本文在 LLM-P2G 模型的基础上，提出一种采样边缘化方法（Sampling-K-Marginalized, SKM），用采样替代束搜索生成多候选音素序列，从而提升音素序列的多样性，改进了边缘化建模。

具体来说，SKM 在音素解码阶段以采样方式替代传统的束搜索，从声学模型输出的概率分布中生成多条候选路径，既缓解了候选路径覆盖范围受限和计算开销较大的问题，又更好近似了边缘化计算。该方法不仅改善了训练收敛速度和稳定性，还在多个多语言语音识别任务中取得了较 TKM 更好识别性能。此外，SKM 在解码资源消耗和推理效率上优于传统 WFST，展示出在资源受限环境中的实际应用潜力。

本文首先介绍了 LLM-P2G 架构及其 Top-K 边缘化（TKM）方法的基本原理，并在此基础上提出采样边缘化（SKM）策略，以提升候选生成的多样性与训练效率。随后，我们在多语言数据集上开展系统实验，验证了 SKM 在训练收敛速度、识别准确率和系统资源消耗方面的优势。为了进一步评估方法的有效性，本文还选取一种常见利用投射器结合大语言模型的语音识别方法（SpeechLLM）作为对比方案，结果表明 SKM 驱动的 LLM-P2G 在保持结构简洁性的同时具备更强的性能表现，进一步验证了其在跨语言语音识别任务中的实用性。

# 1 方法概述

## 1.1 LLM-P2G 架构简介

LLM-P2G（Large Language Model-based Phoneme-to-Grapheme）[6]是一种以音素序列为接口，将声学建模与语言建模解耦的语音识别架构。该方法通过将语音识别任务划分为两个阶段：音频到音素的声学建模阶段与音素到文字的大语言模型建模阶段，实现了模型结构的模块化，便于集成和扩展。其模型结构如图 1 所示：

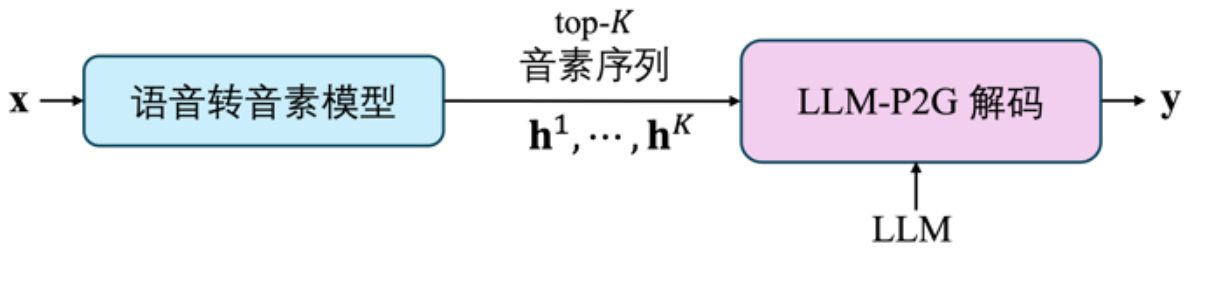

图 1 采用 LLM-P2G 解码的音素语音识别

在第一阶段，输入语音信号经过声学模型解码后输出若干候选音素序列，表示语音到音素（Speech-to-Phoneme，S2P）的预测，基于联结时序分类（CTC）[12]来实现，这一阶段的目标是建模语音信号与音素之间的映射关系。LLM-P2G 的声学模型基于多语言预训练模型 Whistle[10]微调得到。Whistle 采用弱音素监督声学模型[11]，高效支持多语言。

在第二阶段，所得到的音素序列被输入至大语言模型（mT5[1]），进行音素到文字（Phoneme-to-Grapheme, P2G）的转换。该阶段的核心是利用预训练语言模型对目标语言的丰富建模能力，从而提升识别的准确性。由于大语言模型拥有广泛的语言知识，其在语言建模方面的表现远优于传统的有限状态解码器（WFST）。此外，大语言模型在多语言设置下同样具备强泛化能力，有助于支持跨语言场景。

在解码过程中，LLM-P2G 不再依赖传统的 WFST（加权有限状态转移）解码器，而是采用两阶段的音素预测与文字生成方式，使模型架构更为灵活。然而，这种分阶段的设计也引入了新的挑战：声学模型在音素预测过程中可能会产生不确定性，而若仅依赖单一路径进行文字解码，容易造成信息丢失，影响最终识别效果。为此，研究者提出了 Top-K 边缘化（Top-K Marginalized, TKM）[6]方法，通过保留多个优选的音素候选序列，并在训练和解码过程中对这些候选路径进行边缘化计算，有效缓解了信息丢失问题。下面将详细介绍 TKM 的建模原理与实现方式。

## 1.2 Top-K 边缘化（TKM）

受检索增强生成（RAG）[13]技术启发，Top-K 边缘化（Top-K Marginalized, TKM）[6]方法，并不依赖于单一路径作为语言模型的输入，而是在声学模型阶段生成 Top-K 个音素序列候选路径，记作 $\{\mathbf{h}^{(1)}\ \mathbf{h}^{(2)} \ldots \mathbf{h}^{(k)}\}$。对于给定语音特征序列输入 $\mathbf{x}$ 和目标文字序列输出 $\mathbf{y}$，TKM 采用边缘化建模：

$$p(\mathbf{y}|\mathbf{x}) \approx \sum_{\mathbf{h} \in top-K(p(\mathbf{h}|\mathbf{x}))} p(\mathbf{h}|\mathbf{x}) p(\mathbf{y}|\mathbf{h})$$

$$= \sum_{k=1}^{K} p(\mathbf{h}^{(k)}|\mathbf{x}) \prod_{i=1}^{L} p(y_i|\mathbf{h}^{(k)}, y_{1:i-1}) \quad (1)$$

其中，$K$为声学模型 S2P 生成音素时的束搜索宽度，$\mathbf{h}^{(k)}$表示对 S2P 进行束搜索生成的第 $k$ 个音素序列，$p(\mathbf{h}^{(k)}|\mathbf{x})$可以通过 CTC 前向-后向算法进行计算；$L$ 为句子$\mathbf{y} = y_1, \cdots, y_L$的长度，而$p(y_i|\mathbf{h}^{(k)}, y_{1:i-1})$则可以通过基于大语言模型的 P2G 模块以自回归方

式计算得到。TKM先对每个高概率音素路径$\mathbf{h}^{(k)}$分别计算其出现概率与对应的文本生成概率，再整体加权求和，得到最终的输出概率 $p(\mathbf{y}|\mathbf{x})$。

为了提升候选路径的多样性并加快训练收敛速度，LLM-P2G [6]引入了随机化边缘化训练（Randomized-TKM）。在传统的 TKM 中，模型使用束搜索固定生成 Top-K 条音素序列用于边缘化计算，而在 Randomized-TKM 中，进一步引入了一个控制候选选择的超参数 $n$，即从 Top-K 路径中随机采样 $n$ 条音素序列（其中 $n < K$）参与边缘化概率计算。该策略在保证候选质量的同时增强了路径的多样性，有助于提升模型对不同带噪音素序列的鲁棒性，同时降低冗余路径对训练效率的影响。

TKM 的提出使 LLM-P2G 可以充分利用声学模型中的多候选信息，缓解音素接口带来的信息瓶颈问题。相比传统的 WFST 解码方法，TKM 无需构建 WFST，具备更强的可扩展性与语言适应能力，尤其适用于多语言或低资源场景。

### 1.3 采样边缘化（SKM）

尽管 TKM 在性能上已取得显著提升，但其仍存在一定局限性。首先，候选路径由束搜索生成，路径间往往高度相似，覆盖范围有限，影响了边缘化建模。其次，束搜索过程的贪婪特性可能导致训练过程不稳定、收敛速度较慢。因此，本文在 TKM 基础上提出采样边缘化方法（Sampling-K-Marginalized, SKM）。

在 SKM 方法中，依然使用公式（1）进行建模，区别在于声学模型（S2P）不再通过束搜索固定生成 Top-K 条音素路径，也无需引入超参数 n 通过随机选取候选来丰富多样性，而是从基于 CTC 的 S2P 模型采样音素序列候选。在语音特征序列 **x** 的每个位置，我们基于 softmax 分布随机抽取一个符号，来自音素集及空白符。这样，我们就可以从 CTC 模型中采样一条状态路径。然后，将重复音素符号精简为单个符号并移除所有空白符号后，我们从基于 CTC 的 S2P 模型中获得一个样本$\mathbf{h}^{(k)}$（即一个音素序列）。实际操作中，我们会在每个位置独立抽取$K$次，从而获得$K$条独立的状态路径，这些路径最终转换为$K$条音素序列样本。

值得指出的是，SKM 中上述多候选序列是实时采样生成的，进一步提升多样性。SKM 还引入温度采样（Temperature Sampling）：调节温度参数 T，提高低概率音素的选中概率，增强候选多样性。

相比 TKM，SKM 的核心优势在于候选路径分布更广泛、结构差异更显著，从而提供了更充分的音素候选序列。这种方式不仅更好近似了边缘化计算，还在实际训练中展现出更快的收敛速度和更稳定的梯度行为。同时，由于采样机制避免了束搜索的前缀依赖与路径裁剪过程，SKM 在推理资源消耗方面也更具灵活性与效率。

## 2 实验设计

### 2.1 数据集

实验在 Common Voice（CV）[14]数据集（第 11.0 版，2022 年 9 月发布）上进行。我们选取了来自不同语系的两种语言：波兰语（pl）和德语（de），每种语言的训练数据为 130 小时。这两种语言与预训练模型 Whistle 所使用的语言一样，均采用拉丁字母书写形式，并且在预训练的大语言模型（mT5[1]）中也有良好的覆盖。

### 2.2 模型训练与评估

我们的模型训练使用了 CAT [15]语音识别工具包。声学模型采用公开发布的 Whistle-S [10]模型作为骨干网络，该模型是基于 Conformer [11] 的多语言声学模型，使用联结时序分类（CTC）[12] 在 Common Voice 十种语言上进行预训练。

实验基线来自 LLM-P2G 文章[6]中的两个使用 WFST 解码的模型与 LLM-P2G 模型。前两个分别通过微调（Fine-tuning, FT）Whistle-S 骨干模型获得：一个使用弱音素标签进行训练，另一个使用子词标签训练，分别在表 1 中记为“Whistle Phoneme FT” 和 “Whistle Subword FT”，均采用 WFST 解码。最终的词错误率结果，采用了 4-gram 的词级语言模型，在 WFST 解码和束搜索后的重排序阶段均使用该语言模型。这一跨语言实验设置遵循了 Whistle [10] 的方法。

LLM-P2G 是通过在“Whistle Phoneme FT”的 S2P 模型（称为 Whistle-S2P）生成的音素数据上微调 大语言模型 mT5-base [1] 得到的。mT5-base 是一个包含 5.83 亿参数的编码器-解码器结构的 Transformer 模型 [16]，在 mC4 多语言语料（覆盖 101 种语言，包括波兰语和德语）上预训练而成。为了能够进行公平比较，训练过程中，Whistle-S2P 模型始终被冻结，不进行参数更新。

在 LLM-P2G 的 TKM 训练中，超参数 K 和 n 分别设置为 32 和 8。训练得到的 LLM-P2G 模型在

表 1 中被标记为 “LLM-P2G + randomized TKM”。在 TKM 解码阶段，使用 Top-8 的候选路径。而对于本文提出 LLM-P2G 的 SKM 方法，超参数$K$在训练与解码过程中都设置为 8，而温度系数 T 设置为 1.5。

### 2.3 实验结果

为全面评估本文提出的采样边缘化方法（SKM）的有效性，我们将其与传统 WFST 解码方法及基于束搜索的边缘化方法（TKM）进行了性能与显著度比较。所有方法均在相同的声学模型（Whistle-S2P）和语言模型（LLM-P2G）基础上构建，评估指标采用标准的词错误率（Word Error Rate, WER），实验针对波兰语（Polish）和德语（German）的跨语音识别任务进行。实验结果如表 1 和表 2 所示：

表 1 TKM 和 SKM 方法训练的 LLM-P2G 与 WFST 模型的词错误率（WER%）比较

| Model | Polish | German |
|---|---|---|
| Whistle Phoneme FT | 4.30 | 15.73 |
| Whistle Subword FT | 3.82 | 14.01 |
| LLM-P2G-TKM | 3.80 | 13.18 |
| LLM-P2G-randomized TKM | 3.68 | 13.03 |
| LLM-P2G-SKM | **3.61** | **12.94** |

表 2 TKM 方法训练的 LLM-P2G 与 WFST 模型的显著性检验

| Language | p-value |
|---|---|
| Polish (3.68 vs. 3.82) | 1e-04 |
| German (13.03 vs. 14.01) | 7e-23 |

从表 1 与表 2 中可看到如下实验现象：

1) 传统基线模型（Whistle Phoneme FT / Subword FT）与 LLM-P2G 架构相比，在两种语言上均表现出劣势，说明引入大语言模型的两阶段方法在提升语音识别性能方面具有显著效果。
2) 原始 LLM-P2G 方法（TKM）通过 Top-K 候选边缘化显著降低了 WER，特别是在德语中错误率相对减低 17%（15.73% vs 13.03%），证明了多路径建模的有效性。
3) 引入随机路径选择的 randomized-TKM 在两种语言上均优于标准 TKM，说明多样性更高的候选路径有助于进一步缓解候选集中度带来的建模偏差，表 2 的显著性结果表示显著性很强。
4) 本文提出的 SKM 方法在两个语言上进一步降低 WER（波兰语从 3.68% 降至 3.61%，德语从 13.03% 降至 12.94%），验证了采样边缘化建模路径的优势。

### 2.4 解码效率

为评估不同边缘化策略对训练效率的影响，我们对比了 LLM-P2G 模型在采用 TKM 与 SKM 方法下的训练损失下降曲线。训练过程的曲线如图 2 所示。

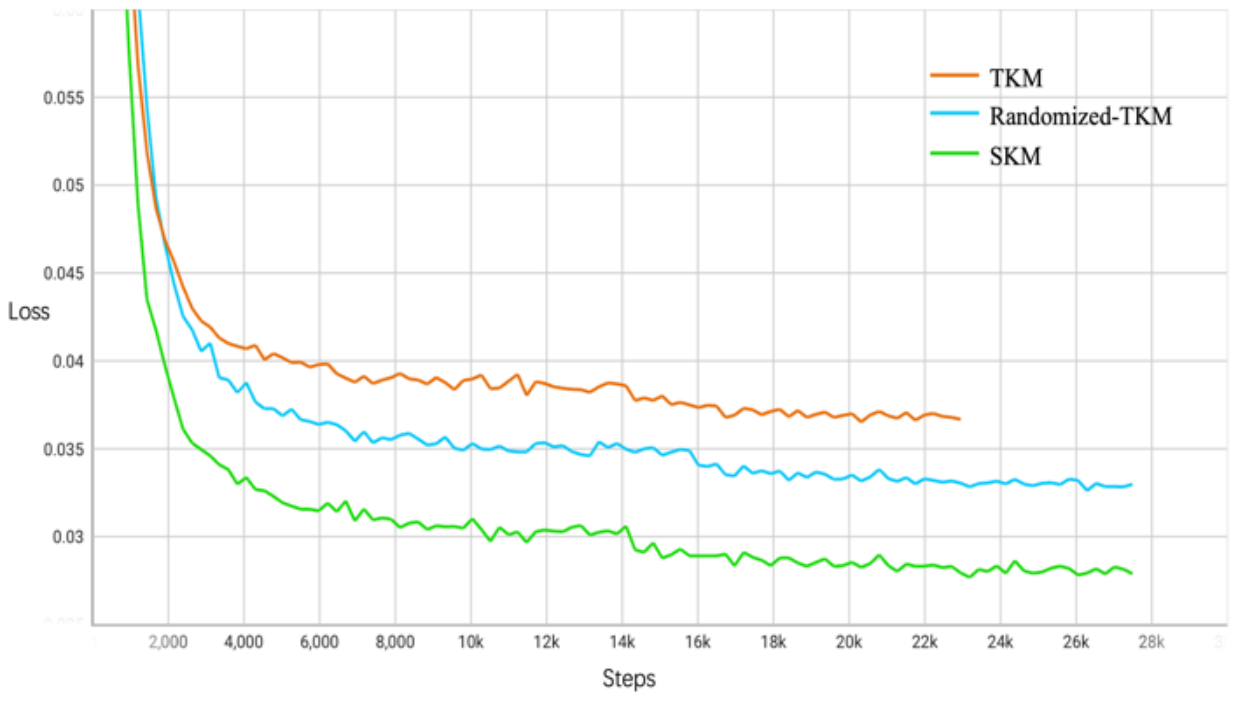

图 2 LLM-P2G 使用 TKM、Randomized- TKM 和 SKM 训练策略时的损失下降曲线对比

从曲线中能看出以下现象：

1) TKM 方法（橙色曲线）：其损失下降速度较慢，且在约 4k 步后趋于平稳，最终收敛在损失值 0.037 附近，说明其边缘化建模过程存在一定限制，难以充分利用训练数据的信息。
2) Randomized-TKM（蓝色曲线）：引入随机性后，模型的收敛速度明显加快，最终稳定在损失值 0.033 左右，说明随机选择候选路径可以缓解候选路径结构单一的问题，提高边缘化建模的有效性。
3) SKM 方法（绿色曲线）：作为本文提出的方法，其损失下降最快，在 3k 步之前即大幅低于其他方法，并最终稳定在 0.027 附近，显著优于 TKM 与 Randomized-TKM。这说明采样生成候选路径相比于束搜索能提供更加多样化的训练信号，

从而提升了模型对复杂输入的建模能力，并改善了训练过程的稳定性与效率。

**2.5 资源消耗**

为评估 LLM-P2G 在实际应用中的资源效率，本节对其与传统 WFST 解码器在计算开销和存储占用等方面进行了对比分析。尽管 LLM-P2G-TKM 在准确率上已展现出优越性能，但其引入了大语言模型，可能导致额外的计算开销。因此，我们从多个维度对比两种模型的资源消耗情况，包括 CPU 内存占用、GPU 显存使用、模型文件存储空间以及解码实时系数（Real-Time Factor, RTF），以评估 LLM-P2G 在推理效率和部署成本方面的可接受性，并验证其是否能够在替代 WFST 的同时维持良好的运行效率。

值得注意的是，相比于 TKM 方法，SKM 的改进只针对模型训练，而对于解码过程，依然遵循 TKM 方法，即解码时始终使用前 K=8 个候选用于边缘化计算；此外 LLM-P2G 的另一方法噪声数据增强（Data Augmentation with Noisy Phonemes, DANP）[6]也进行了比较。LLM-P2G 模型与 WFST 模型的资源消耗情况如表 3 所示：

表 3 TKM 方法训练的波兰语 LLM-P2G 模型与 WFST 模型解码时的资源消耗比较

| Model | CPU | GPU | Storage | RTF |
|---|---|---|---|---|
| Whistle Phoneme FT | **4.0 GB** | **1.5 GB** | **0.7 GB** | **0.02** |
| Whistle Subword FT | 5.9 GB | **1.5 GB** | 2.4 GB | 0.07 |
| LLM-P2G + DANP | **4.0 GB** | 4.6 GB | 2.5 GB | 0.07 |
| LLM-P2G + TKM/SKM | **4.0 GB** | 6.3 GB | 2.5 GB | 0.10 |

实验现象如下：

1) RTF（实时因子）对比：LLM-P2G + DANP 的 RTF 为 0.07，与传统的 Whistle Subword FT 持平，意味着加入大模型后的解码效率仍处于可接受范围，并没有带来显著损耗。LLM-P2G + TKM 由于涉及多路径边缘化计算，RTF 上升至 0.10，但仍可用于大多数应用场景。
2) GPU 显存对比：LLM-P2G + DANP 约为 4.6 GB，TKM 上升到 6.3 GB，相比传统模型（1.57 GB）明显上升。这反映了使用 LLM（如 mT5）在 P2G 阶段对 GPU 资源的需求更高，但仍在主流显卡（如 8GB~16GB）的处理能力范围内。
3) CPU 内存以及存储对比：LLM-P2G 模型 CPU 占用和 Whistle phoneme FT 持平（4.0 GB），低于子词微调(Whistle Subword FT)模型（5.9 GB）。存储空间消耗略高（2.5 GB），但相比子词 WFST 模型没有明显劣势。

**2.6 与 SpeechLLM 的性能对比**

为进一步验证本文提出的 SKM 方法驱动的 LLM-P2G 框架在语音识别任务中的有效性，我们选取当前有代表性的语音识别与大语言模型融合方案 —— SpeechLLM [17] 作为对比对象。SpeechLLM 通过适配器（Adapter）将声学神经网络的输出特征向量投射接入语言模型，然后完成文本生成。我们实验中采用池化适配器（Pooling Adapter），它由平均池化层、线性投影层和两层含 GELU 激活的多层感知机组成，参数约 2 MB。

为确保对比公平性，我们在 SpeechLLM 方法中采用了 LLM-P2G 所使用的 Whistle 声学模型与 mT5 大语言模型，命名为 “Whistle + pooling-adapter + mT5”。此外，我们还将 LLM-P2G 原始方法（DANP 和 TKM）加入对比，进一步分析两阶段结构与端到端结构的性能差异。

LLM-P2G 模型与 SpeechLLM 模型的性能比较如表 4 所示。

表 4 LLM-P2G 与 SpeechLLM 在波兰语（Polish）和德语（German）上的词错误率（WER%）比较

| Model | Polish | German |
|---|---|---|
| Whistle + pooling-adapter + mT5 | 4.15 | 13.35 |
| LLM-P2G + DANP | 4.18 | 13.63 |
| LLM-P2G + randomized TKM | 3.68 | 13.03 |
| LLM-P2G + SKM | **3.61** | **12.94** |

对比结果表明：

1) SpeechLLM 方法略优于 DANP 驱动的 LLM-P2G，说明在适配器投射结构下，大语言模型可以较好地直接建模声学特征，具备一定端到端建模优势。
2) TKM 或 SKM 驱动的 LLM-P2G 模型在 Polish 与 German 上均优于 SpeechLLM，

分别实现了约 13% 与 3% 的相对词错误率下降，达到当前任务的最优性能（SOTA）。

该对比实验表明，显式音素建模提供了结构上的清晰分层，而 TKM 增强了候选路径的建模能力，从而更有效地发挥大语言模型的上下文建模优势。相比端到端的结构更具灵活性和扩展性；此外，SKM 中的采样策略缓解了束搜索路径间的同质性问题，提升了训练效率与候选序列质量，进一步彰显了边缘化建模方法在多语言语音识别系统中的价值与潜力。

## 3 结论

本文围绕两阶段语音识别框架 LLM-P2G 中 Top-K 边缘化方法（TKM）存在的候选多样性不足与训练效率低的问题，提出了一种基于采样的改进策略 —— 采样边缘化方法（SKM），并在跨语言识别任务中进行了系统评估。研究结果表明，在与已有的 TKM 和 randomized-TKM 方法的比较中，采样机制在候选音素路径生成阶段引入了更大的随机性和多样性，缓解了传统束搜索方法中由于贪婪性导致的路径同质化问题，从而改进了边缘化建模，提升了训练收敛速度。此外，本文还进行了与一种当前常用的结合大语言模型的语音识别方案 SpeechLLM 的对比实验。结果显示， LLM-P2G 在不增加模型复杂度的情况下，在识别性能上取得显著优势。总的来说，本文验证了大语言模型和采样边缘化驱动的基于音素的语音识别的可行性与优势。以音素连接语音与语言，并结合大语言模型的新框架的更多潜力，值得进一步探索。

## 参考文献（References）

# Phoneme-based speech recognition driven by large language models and sampling marginalization

Ma Te[1], Li Nanjie[1], Huang Hao[1], Ou Zhijian[2]
(1. School of Computer Science and Technology, Xinjiang University, Urumqi, 830046, China;
2. Speech Processing and Machine Intelligence (SPMI) Lab, Tsinghua University, Beijing, 100084, China)

**Abstract:** Recently, the Large Language Model-based Phoneme-to-Grapheme (LLM-P2G) method has shown excellent performance in speech recognition tasks and has become a feasible direction to replace the traditional WFST decoding method. This framework takes into account both recognition accuracy and system scalability through two-stage modeling of phoneme prediction and text generation. However, the existing LLM-P2G adopts the Top-K Marginalized (TKM) training strategy, and its candidate phoneme sequences rely on beam search generation, which has problems such as insufficient path diversity, low training efficiency, and high resource overhead. To this end, this paper proposes a sampling marginalized training strategy (Sampling-K Marginalized, SKM), which replaces beam search with random sampling to generate candidate paths, improving marginalized modeling and training efficiency. Experiments were conducted on Polish and German datasets, and the results showed that SKM further improved the model learning convergence speed and recognition performance while maintaining the complexity of the model. Comparative experiments with a speech recognition method that uses a projector combined with a large language model (SpeechLLM) also show that the SKM-driven LLM-P2G has more advantages in recognition accuracy and structural simplicity. The study verified the practical value and application potential of this method in cross-language speech recognition systems.